\documentclass[%
 reprint,
 amsmath,amssymb,
pra,
]{revtex4-1}

\usepackage{graphicx}
\usepackage{dcolumn}
\usepackage{bm}

\graphicspath{%
    {converted_graphics/}
    {/}
}
\begin{document}


\title{Inductively Coupled Augmented Railgun}

\author{Thomas B. Bahder}  \author{William C. McCorkle}%
\affiliation{%
Aviation and Missile Research, 
      Development, and Engineering Center, \\    
US Army RDECOM, 
Redstone Arsenal, AL 35898, 
U.S.A.}%

\date{\today}
  
\begin{abstract}
We derive the non-linear dynamical equations for an augmented electromagnetic railgun, whose augmentation circuit is inductively coupled to the gun circuit.  We solve these differential equations numerically using example parameter values.  We find a complicated interaction between the augmentation circuit, gun circuit, and mechanical degrees of freedom, leading to a  complicated optimization problem. For certain values of parameters, we find that an augmented electromagnetic railgun has an armature kinetic energy that is 42\% larger than the same  railgun with no augmentation circuit. Optimizing the parameters may lead to further increase in performance.
\end{abstract}

\maketitle


\section{\label{Introduction}Introduction}
Launch systems, such as the electromagnetic railgun (EMG), have complex electrical transient phenomena~\cite{14th_Symposium_2008,McCorkle2008}. During launch, the transient involves the build-up and penetration of a magnetic field into the metallic material.  The dynamics of magnetic field penetration into the metal rails is described by a well-known magnetic diffusion equation \cite{Knoepfel2000}, which leads to a skin effect  where a large current is transported inside a narrow channel.   In a railgun that has a moving conducting armature, the effect is called a velocity skin effect (VSE) and is believed to be one of the major problems in limiting railgun performance~\cite{Knoepfel2000,Young1982}, because it leads to intense Joule heating of the conducting materials, such as rails and armatures. To what extent the VSE  is responsible for limiting the performance of solid armatures is still the subject of research~\cite{Drobyshevski1999,Stefani2005,Schneider2007,Schneider2009}.  

For high-performance EMGs, such as those planned by the navy for nuclear and conventional warships~\cite{Walls1999,BlackThesis2006,mcnab2007}, the armature velocity must be increased while keeping the length of the rails fixed. One approach to increasing the armature velocity is to use some sort of augmentation to the EMG circuit.  Various types of augmentation circuits have been considered~\cite{Kotas1986}, including hard magnet augmentation fields~\cite{Harold1994} and superconducting coils~\cite{Homan1984,Homan1986}. 

In this paper, we consider a lumped circuit model of an augmented EMG. The augmented EMG consists of two inductively coupled circuits.  The first circuit is the EMG circuit, containing the rails connected to a voltage source, $V_g(t)$, that powers the rails and armature.  The augmentation consists of an external circuit that is  inductively coupled to the EMG circuit via magnetic field, but is not connected electrically to the EMG circuit.  The augmentation circuit has its own voltage generator, $V_a(t)$.  See Figure 1 for a schematic layout.  Figure 2 shows the equivalent lumped-circuit model that we are considering.  

A simplified arrangement has been previously considered, where the EMG circuit was augmented by a constant external magnetic field~\cite{Harold1994}.  In our work, we assume that a real augmentation circuit produces the magnetic field  that couples to the EMG circuit, and hence, the EMG circuit is interacting with the augmentation circuit through mutual inductance, see Figure 1.  This results in a ``back action" on the augmentation circuit by the EMG circuit. This back action modifies the current in the augmentation circuit, resulting in a non-constant $B$-field due to the augmentation circuit.   Furthermore, the EMG circuit is coupled to the mechanical degree of freedom, the moving armature, which leads to a variation (with armature position) of the self inductance and resistance of the EMG circuit.   The variations  of the self inductance, resistance and mutual inductance in the EMG circuit  act back on the augmentation circuit, leading to a complex interaction between the gun circuit, the augmentation circuit, and the mechanical degree of freedom (the armature). The resulting dynamical system is described by three non-linear differential equations that are derived in Section~\ref{DynamicalEquations}.  In order to derive these dynamical equations, we first consider Ohm's law in the presence of an external magnetic field, which is the field produced by the augmentation circuit.

\begin{figure}[tbp] 
  \centering
   \includegraphics[width=3.2in]{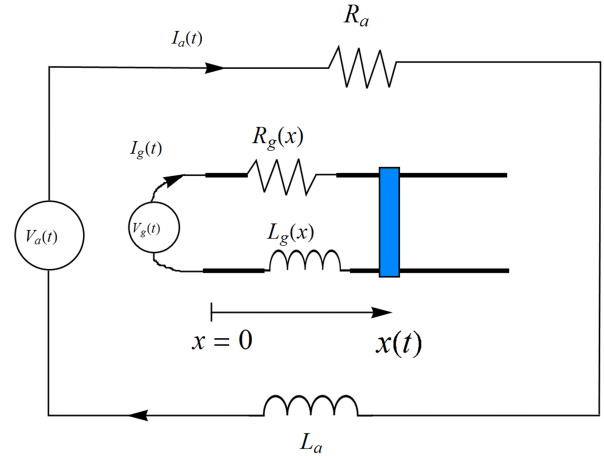}
  \caption{Schematic diagram of an inductively augmented EMG with its surrounding augmentation circuit.  Magnetic flux from the augmentation circuit inductively links to the gun circuit. 
  \label{fig:AugmentedEMG}}
\end{figure}

\section{\label{OhmsLaw}Ohm's Law in a Magnetic Field}
Consider a series electrical circuit with a voltage generator, $V$, resistor $R$, and inductor $L$.  Assume that an external source of current creates a time-dependent magnetic field, 
${\bf B}_e$, and at some time a quantity of magnetic flux from this field, $\phi_e$,  passes through the electrical circuit.  The time-dependent external magnetic field, ${\bf B}_e$, creates a non-conservative electric field, ${\bf E}_e$, which are related by the Maxwell equation
\begin{equation}
{\rm curl}  \,{\bf{E}}_e  =  - \frac{\partial {\bf B}_e }{\partial t}
\label{MaxwellEq} 
\end{equation}
Taking an integration contour along the wire of the electrical circuit, and integrating Eq.~(\ref{MaxwellEq}) over the area enclosed by this contour, using Stoke's rule to change the surface integral to a line integral, we have the emf induced in the electrical circuit, $\varepsilon_e$, given in terms of the rate of change of the external flux (due to the external magnetic field
${\bf B}_e$):
\begin{equation}
\varepsilon _e  = \oint {{\bf{E}}_e  \cdot d{\bf{l}}}  =  - \frac{{\partial \phi _e }}{{\partial t}}
\label{emf} 
\end{equation}
There are two sources that do work on the electrical circuit, the voltage generator, $V$, and the emf, $\varepsilon_e$, induced in the circuit by the external magnetic field, ${\bf B}_e$.  The power entering  the circuit, $\varepsilon_e \, I + I \, V$, goes into Joule heating in the resistor and into increasing the stored magnetic field energy: 
\begin{equation}
\varepsilon_e \, I + I \, V = I^2 R + \frac{d}{dt} \frac{1}{2} L I^2 
\label{power} 
\end{equation}
Dividing Eq.~(\ref{power}) by $I$, and solving for the externally induced emf, $\varepsilon_e$, in terms of the circuit quantities and the rate of change of external flux,  $\phi _e$,
\begin{equation}
\varepsilon_e = -V  + I R  + L  \frac{d I}{dt} =  - \frac{d \phi_e}{dt} 
\label{emfTotal} 
\end{equation}
Equation~(\ref{emfTotal}) can be rewritten in terms of the total emf as
\begin{equation}
-V  + I R  =  - L  \frac{d I}{dt}   - \frac{d \phi_e}{dt} \equiv  - \frac{d \phi}{dt}
\label{emfTotal2} 
\end{equation} 
where $- L  {d I}/{dt} = {d \phi_i}/{dt}$ is the emf generated in the circuit due to the internal current $I$ in the circuit, and $- {d \phi_e}/{dt}$ is the emf generated in the circuit by the external magnetic field 
${\bf B}_e$.   The sum of the internal and external flux  is the total flux: $\phi = \phi_i + \phi_e$.   Equation~(\ref{emfTotal2}) relates the rate of change of total flux through the circuit, $ d \phi / d t $, to the voltage drops along the circuit, $-V + I R$, and is an expression of Ohm's law in the presence of a magnetic field.  We will use this form of Ohm's law to write the coupled equations for the augmented railgun.

\section{\label{DynamicalEquations}Dynamical Equations}

\begin{figure}[tbp] 
  \centering
   \includegraphics[width=3.2in]{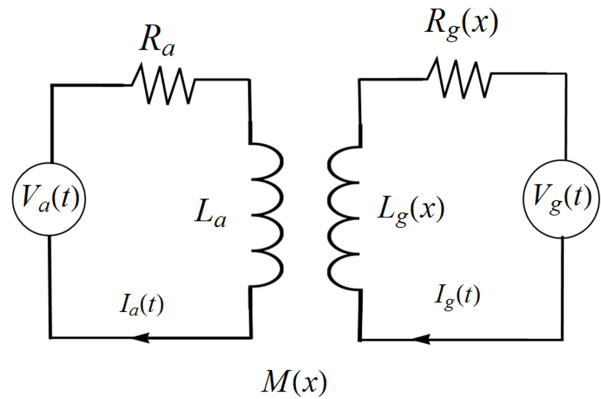}
  \caption{The equivalent circuit is shown for the augmented railgun in Figure~\ref{fig:AugmentedEMG}.   Magnetic flux from the augmentation circuit couples to the gun circuit through mutual inductance $M(x)$.  The self inductance of the gun circuit, $L_g(x)$,  the mutual inductance, $M(x)$, and the resistance, $R_g(x)$,  are functions of the armature position, $x(t)$.  
  \label{fig:EquivalentCircuit}}
\end{figure}

Consider an augmented railgun composed of an augmentation circuit with voltage generator $V_a(t)$ and a gun circuit with voltage generator, $V_g(t)$.  We assume that the circuits are inductively coupled, but have no  electrical connection, see Figure~\ref{fig:AugmentedEMG}.  The equivalent circuit for the augmented railgun is shown in Figure~\ref{fig:EquivalentCircuit}.  The resistance of the gun circuit, $R_g(x)$, changes with armature position $x(t)$ and can be written as
\begin{equation}
R_g(x)= R_{g0} + R_g^\prime \, \,  x(t)
\label{SelfInductanceGun} 
\end{equation}
where $R_{g0}$ is the resistance of the gun circuit when $x=0$, and $R_g^\prime$ is the gradient of resistance of the gun circuit at $x=0$.

The total flux in the gun circuit, $\phi_g$, and the total flux in the  augmentation circuit, $\phi_a$, can be written as
\begin{eqnarray}
\phi _g & = & L_g I_g + M_{\text{ga}}I_a \\ 
\phi _a & = & L_a I_a + M_{\text{ag}}I_g
\label{totalflux}
\end{eqnarray}
where $I_a$ and $I_g$ are the currents in the augmentation circuit and gun circuit, respectively,  $L_a$ and $L_g$, are the self inductances of the  augmentation and gun circuits, respectively, and $M_{\text{ga}}$ and $M_{\text{ag}}$ are the mutual inductances, which must be equal, $M_{\text{ga}} = M_{\text{ag}} =M(x)$. The self inductance of the gun circuit, $L_g(x)$, changes with armature position $x(t)$.  Also, the area enclosed by the gun circuit changes with armature position, and therefore, the coupling between the augmented circuit and gun circuit, represented by the mutual inductance, $M(x)$, changes with armature position, $x(t)$.  Furthermore, in order for the free energy of the system to be positive, the self inductances and the mutual inductance must satisfy~\cite{LL_continuous_media}
\begin{equation}
M(x)=k\sqrt{L_a L_g(x)}
\label{InductanceIdentity} 
\end{equation}
for all values of $x$.  Here, the coupling coefficient must satisfy $|k| < 1 $.   We can write the self inductance of the gun circuit as 
\begin{equation}
L_g(x)= L_{g0} + L_g^\prime \, \,  x(t)
\label{SelfInductanceGun} 
\end{equation}
where $L_{g0}$ is the inductance when $x=0$ and  $L_g^\prime$ is the inductance gradient of the gun circuit.  Similarly, the mutual inductance between augmented circuit and gun circuit can be written as
\begin{equation}
M(x)= M_0 + M^\prime \,\, x(t)
\label{MutualInductance} 
\end{equation}
where $M_0$ is the mutual inductance when $x=0$ and $M^\prime $ is the mutual inductance gradient.  For $x=0$ and $x=\ell$, where $\ell$ is the rail length, Eq.~(\ref{InductanceIdentity}) gives 
\begin{eqnarray}
 M_0 & =  & k\sqrt{L_a L_{\text{g0}}} \label{MutualInductanceM0} \\
M^\prime & = & \frac{1}{\ell } \left[k \sqrt{L_a \left(L_{\text{g0}}+L_g' \, \ell \right)}-M_0\right] \label{MutualInductanceMP}
\end{eqnarray}

The coupling coefficient, $k$, can be positive or negative, and as mentioned above, must satisfy $|k| < 1 $.  The sign of $k$ determines the phase of the inductive coupling between augmentation and gun circuits.   Choosing the  coupling coefficient $k$, and the two self inductances, $L_{g0}$ and $L_{a}$,   Eq.~(\ref{MutualInductanceM0}) then determines the value of the mutual inductance, $M_0$.  Then, choosing  a value for the rail  length  $\ell$, and the self inductance gradient, $L_g^\prime$, Eq.~(\ref{MutualInductanceMP}) determines the mutual inductance gradient,  $M^\prime$. See Table~\ref{gun_parameters} for parameter values used. 
\begin{table*}
\caption{\label{gun_parameters}Electromagnetic gun and augmentation circuit parameters.}
\begin{ruledtabular}
\begin{tabular}{lcl}
Quantity & Symbol &  Value \\
\hline
length of rails (gun length) &  $\ell$  &  10.0 m \\
mass of armature  & $m$ &  20 kg  \\
coupling coefficient   &  $k$  &  $\pm$0.80  \\ 
self inductance of rails at $x=0$  &  $L_{g0}$   &  6.0$\times$10$^{-5}$ H  \\
self inductance of augmentation circuit  &  $L_{a}$   &  6.0$\times$10$^{-5}$ H  \\
self inductance gradient of rails  &  $L_g^\prime$   &  0.60$\times$10$^{-6}$ H/m  \\
resistance of augmentation circuit  &  $R_{a}$   &  0.10 $\Omega$  \\
resistance of gun circuit at $x=0$   &  $R_{g0}$   &  0.10 $\Omega$  \\
resistance gradient of gun circuit at $x=0$  &  $R_g^\prime$   &  0.002 $\Omega/$m  \\
voltage generator amplitude in gun circuit   &  $V_{g0}$   &  8.0$\times$10$^{5}$ Volt   \\
voltage generator amplitude in augmentation circuit   &  $V_{a0}$   &  8.0$\times$10$^{5}$ Volt \\
\end{tabular}
\end{ruledtabular}
\end{table*}

Two dynamical equations for the augmented railgun are obtained by applying Ohm's law in Eq.~(\ref{emfTotal2}) to the gun circuit and to the augmented circuit:  
\begin{eqnarray}
-V_a + I_a R_a  & = &  - \frac{d}{d t} \left[  L_a I_a + (M_0 + M^\prime \, x(t)) \, I_g \right]  \nonumber  \\
-V_g + I_g R_g(x)  & = &    - \frac{d}{d t} \left[  (L_{g0} + L_g^\prime \, x(t)) \, I_g     \right.  \nonumber \\  
      &  & + \left.  (M_0 + M^\prime \, x(t)) \, I_a \right] 
\label{circuitEquations}
\end{eqnarray}
The third dynamical equation is obtained from the coupling of the electrical and mechanical degrees of freedom\cite{McCorkle2008}.  Therefore, the three non-linear coupled dynamical equations for  $I_g(t)$, $I_a(t)$ and $x(t)$ are given by:
\begin{widetext}
\begin{eqnarray}
 -V_a(t)+I_a(t) R_a & = &  -L_a\frac{dI_a}{dt}-\left(M_0+M' x(t) \right)\frac{dI_g}{dt}-M'I_g(t)\frac{dx(t)}{dt} \label{dynamicEq1} \\
 -V_g(t)+I_g(t) \left(R_{\text{g0}}+R_g' \, x(t) \right) & = & -\left(L_{\text{g0}} + L_g' \, x(t) \right) \frac{dI_g}{dt}-L_g' \, \text{  }I_g(t)\frac{dx(t)}{dt}-\left(M_0+M' x(t) \right)\frac{dI_a}{dt}-M' \, I_a(t) \frac{dx(t)}{dt}          \hspace{0.25in}      \label{dynamicEq2}  \\
 m\frac{d^2x(t)}{dt^2}  &  =  & \frac{1}{2}L_g' \,\, I_g{}^2(t)   \label{dynamicEq3}
\end{eqnarray}
\end{widetext}
where $V_g(t)$ and $V_a(t)$ are the voltage generators that drive the gun and augmentation circuits.    From Eq.~(\ref{dynamicEq3}), we see that the EMG armature has a positive acceleration independent of whether the gun voltage generator is a.c. or d.c. because the armature acceleration is proportional to $I_g{}^2(t)$.  The armature velocity is essentially the integral of $I_g{}^2(t)$, and therefore a higher final velocity will be achieved for d.c. current, and associated d.c. gun voltage, $V_g(t)=V_{g0}$, where $V_{g0}$ is a constant.  Of course, the actual current will not be constant in the gun circuit because of the coupling to the moving armature and to the augmentation circuit.   We want to search for solutions where the armature velocity is higher for a EMG with the augmented circuit than for an EMG without an augmentation circuit.  In order to increase the coupling between the augmentation circuit and the gun circuit, we choose an a.c. voltage generator in the augmentation circuit:
\begin{equation}
V_a(t)= V_{a0} \sin ( \omega t)
\label{augmentationCircuitVoltage} 
\end{equation}
where $V_{a0}$ is a constant amplitude for the voltage generator and $\omega$ is the angular frequency of the augmentation circuit voltage generator. 

As an example of the complicated coupling between augmentation circuit and gun circuits, we will also obtain solutions for a d.c. voltage generator for the augmentation circuit.  As we will see, the gun circuit causes a back action on the augmentation circuit, leading to a non-constant current in the augmentation circuit.

We need to choose initial conditions at time $t=0$.  We assume that there is no initial current in the gun and augmentation circuits and that the initial position and velocity of the armature are zero:
 \begin{eqnarray}
 I_g(0) & =  & 0 \label{initial_1} \\
 I_a(0) & =  & 0 \label{initial_2} \\
   x(0) & =  & 0 \label{initial_3} \\
  \frac{d x(0)}{d t} & =  & 0 \label{initial_4} 
\end{eqnarray} 

\begin{figure*}[tbp] 
  \centering
   \includegraphics[width=6.5in]{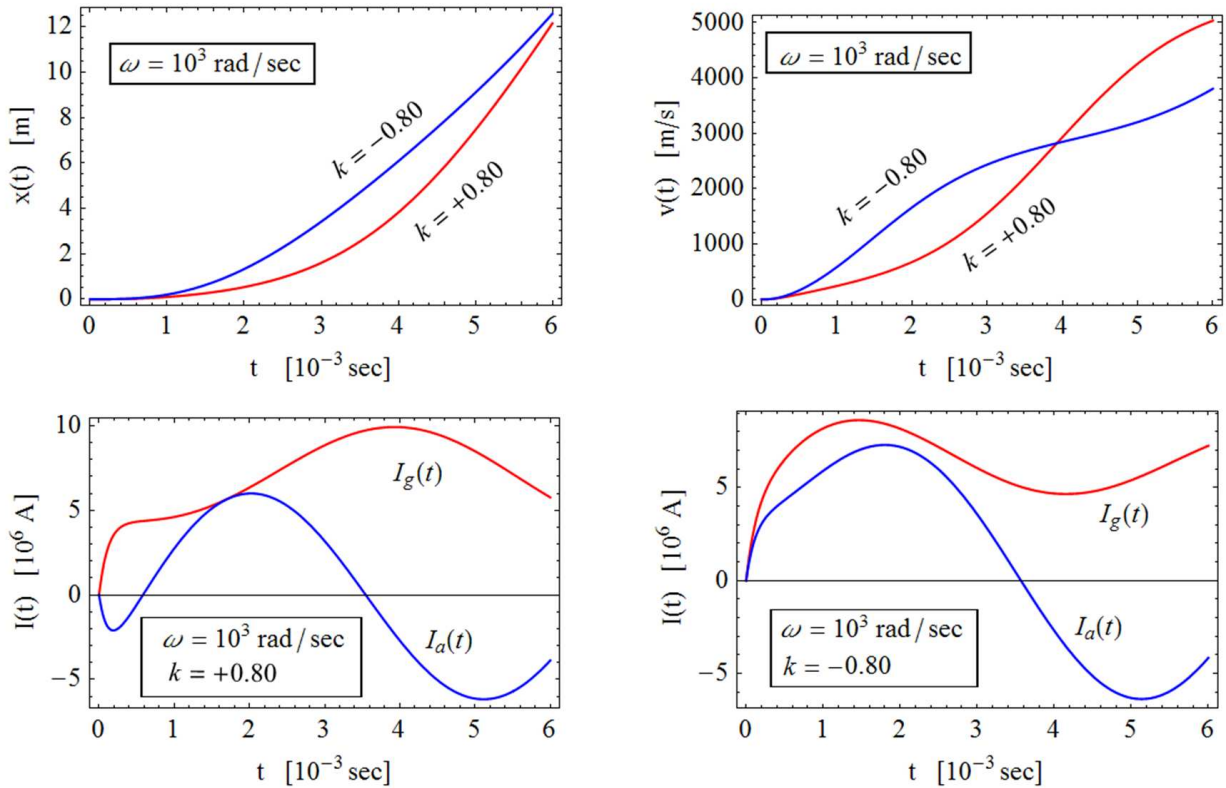}
  \caption{For a.c. augmentation circuit voltage, with $\omega=10^3$ rad/sec, the armature position, velocity, and augmentation and gun circuit currents are plotted for two different values of coupling constant, $k=+0.80$ and $k=-0.80$.  
  \label{fig:CircuitKDependence}}
\end{figure*}
\begin{figure*}[tbp] 
  \centering
   \includegraphics[width=6.9in]{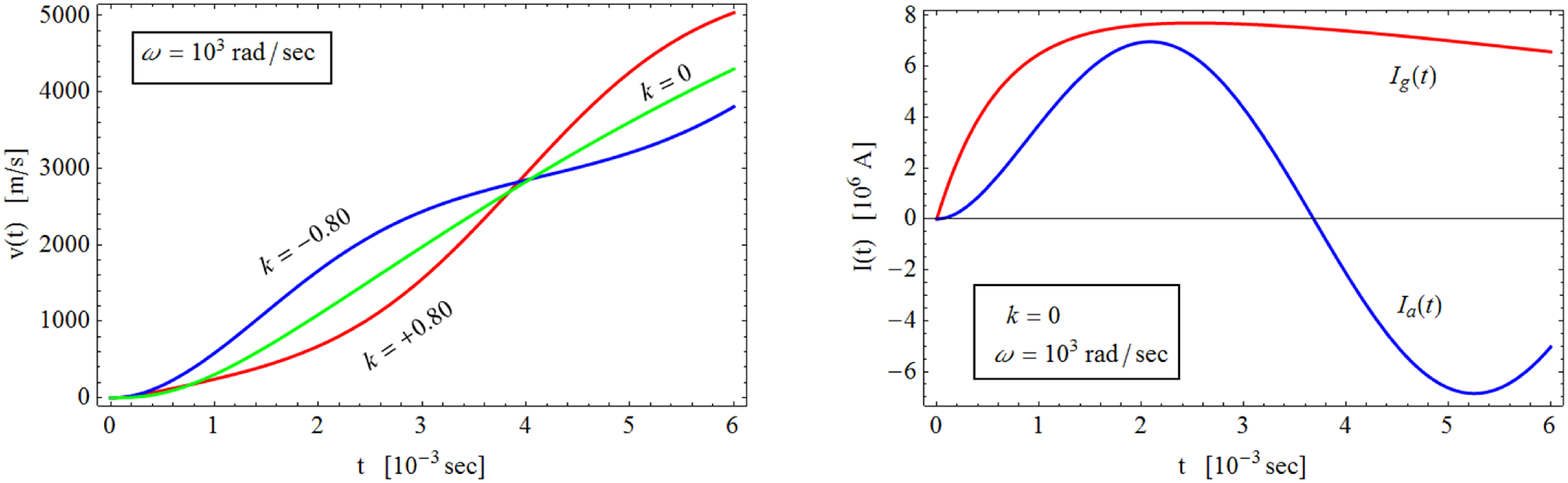}
  \caption{For a.c. augmentation circuit voltage, with $\omega=10^3$ rad/sec, the left plot shows the armature velocity vs. time for three values of coupling constant, $k=-0.80$, $k=+0.80$, and $k=0.0$, for parameter values in Table~\ref{gun_parameters}. Note that the velocity curves in this figure for $k=-0.80$ and $k=+0.80$, are the same as the curves in Figure~\ref{fig:CircuitKDependence}.   The right side plot shows the current in the gun circuit and in the decoupled ($k=0$) augmentation circuit.   The corresponding kinetic energy plots are shown in Figure~\ref{KEplot}.  
  \label{fig:k0}}
\end{figure*}
For the special case when $L_g^\prime=0$, $M^\prime=0$, and $R_g^\prime=0$, the mechanical degree of freedom  described by Eq.~(\ref{dynamicEq3}) decouples from  Eqs.~(\ref{dynamicEq1})--(\ref{dynamicEq3}).  In this case, Eqs~(\ref{dynamicEq1})--(\ref{dynamicEq2}) describe a transformer with primary and secondary circuits having voltage generators, $V_a(t)$ and $V_g(t)$, respectively.  The solution for the mechanical degree of freedom is then $x(t)=0$ for all time $t$.   

In what follows, we solve the dynamical Eqs.~(\ref{dynamicEq1})--(\ref{dynamicEq3}) numerically for the case when the EMG and augmented circuits are coupled.
\begin{figure}[tbp] 
  \centering
   \includegraphics[width=3.5in]{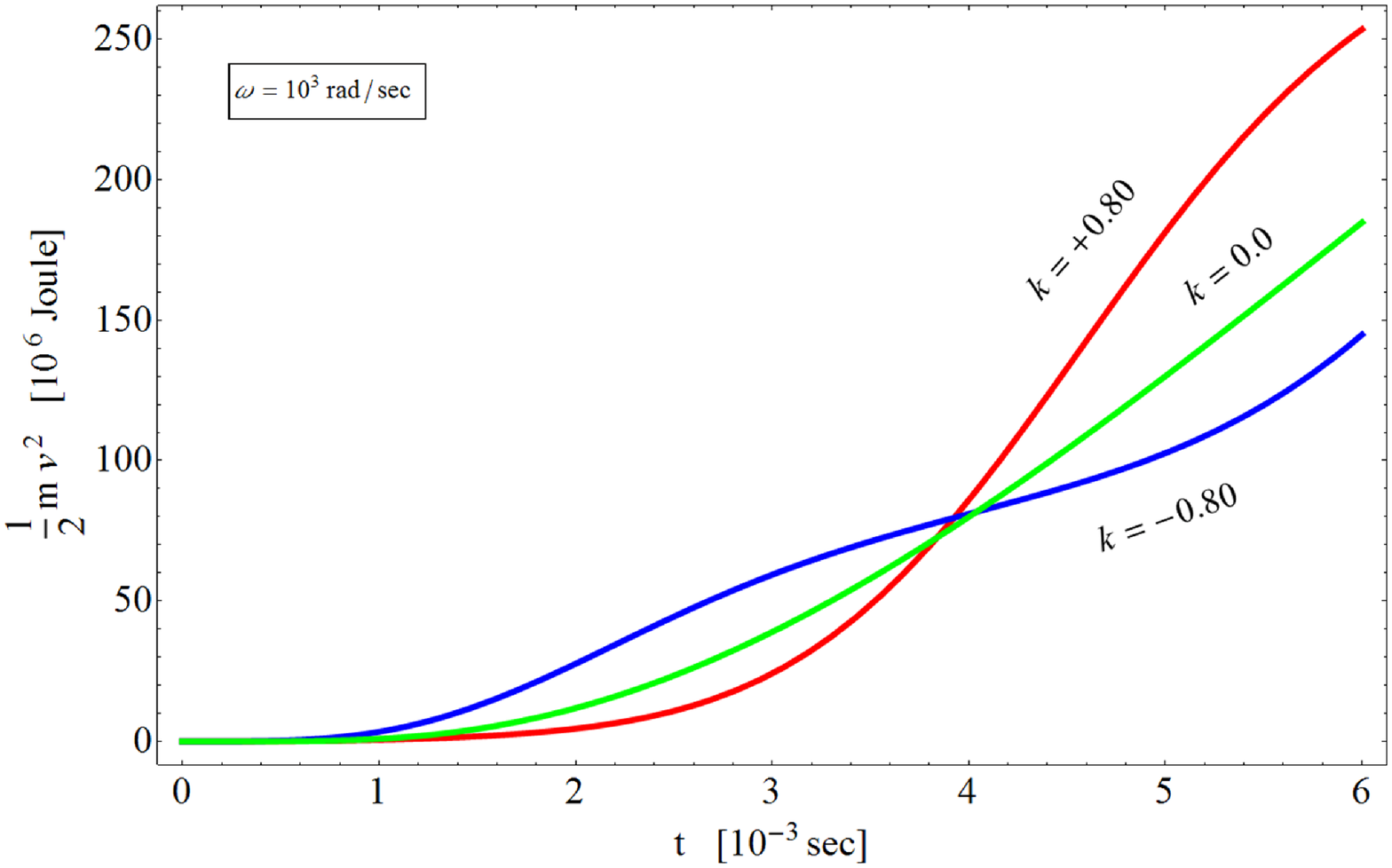}
  \caption{For a.c. augmentation circuit voltage, with $\omega=10^3$ rad/sec, the armature kinetic energy is plotted vs. time for three values of coupling coefficient, $k=-0.80$, 0.0, and 0.80, for parameters values shown in Table~\ref{gun_parameters}.  The kinetic energy is largest when $k=+0.80$.  For the corresponding velocity plots, see Figure~\ref{fig:k0}.    
  \label{KEplot}}
\end{figure}
\begin{figure*}[tbp] 
  \centering
   \includegraphics[width=6.9in]{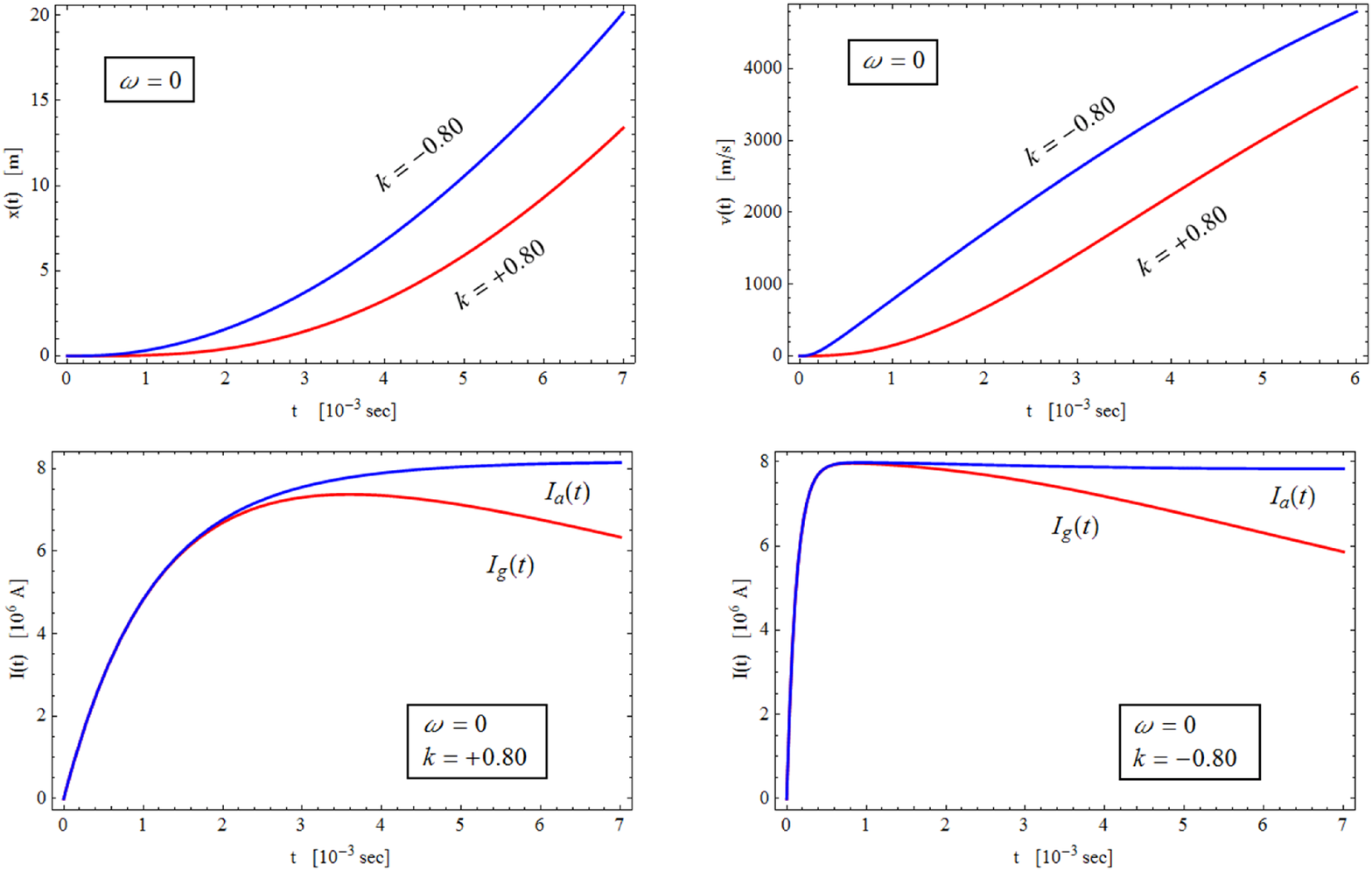}
  \caption{For d.c. augmentation circuit voltage (with $\omega=0$), the armature position, velocity, and currents are shown for the case when the voltage generators in augmentation circuit, $V_a(t)$, and gun circuits, $V_g(t)$, are d.c. generators ($\omega=0$) having values shown in Table~\ref{gun_parameters}.      
  \label{fig:w0}}
\end{figure*}

\section{\label{NumericalSolution}Numerical Solution}
As described above, if we choose the parameters $k$, $L_a$, $L_{g0}$, then $M_0$ is determined from Eq.~(\ref{MutualInductanceM0}).  Next, if we choose $L_g^\prime$ and $\ell$, then the inductance gradient, $M^\prime$, is determined  by Eq.~(\ref{MutualInductanceMP}). See Table~\ref{gun_parameters} for values of parameters used in the calculations below.

The sign of the coupling coefficient, $k$, affects the interaction of the augmentation and gun circuits in subtle ways. We solve Eqs.~(\ref{dynamicEq1})--(\ref{dynamicEq3}) for a positive and negative value of the coupling coefficient $k$.     Figure~\ref{fig:CircuitKDependence} shows a plot of armature position, $x(t)$, velocity $v(t)$, and currents $I_g(t)$ and $I_a(t)$, for a gun circuit powered by a d.c. voltage $V_{g0}$, and an augmentation circuit with an a.c. voltage source given by Eq.(\ref{augmentationCircuitVoltage}), using the parameters in Table~\ref{gun_parameters}  and coupling coefficient values  $k=+0.80$ and $k=-0.80$.  Note that the rail length is taken to be $\ell=10$ m, so the plots are only valid for time $0 \le t \le t_f$, where $t_f$ is given by $x(t_f)=\ell$.  
For $k=+0.80$  we have $t_f=5.563$ ms,  and for $k=-0.80$ we have  $t_f=5.277$ ms, see Table~\ref{gun_performance}.  From Figure~\ref{fig:CircuitKDependence}, for $k=+0.80$, for short times  $t\approx 2$ ms the armature velocity is smaller than for $k=-0.80$.   However, for longer times $t\approx 5$ ms, the armature velocity is higher for $k=+0.80$.  For $k=+0.80$, for short time $t\approx 2$ ms, the augmentation circuit current $I_a(t)$ is out of phase with the gun current $I_g(t)$.   However, for $k=-0.80$,  for short time $t\approx 2$ ms, the augmentation circuit current and gun circuit currents are approximately in phase. Even though the gun circuit has a d.c. voltage generator, the gun circuit current varies due to the coupling with the augmentation circuit and with the mechanical degree of freedom $x(t)$.  These plots illustrate the complicated coupling between the augmentation circuit, the gun circuit, and the mechanical degree of freedom, $x(t)$. When the armature reaches the end of the rails at $\ell=10$ m, its velocity is $v(t_f)=4769$ m/s for $k=+0.80$, whereas its velocity is $v(t_f)=3336$ m/s for $k=-0.80$. The armature kinetic energy  for $k=0.0$ is 159.6 MJ, whereas for $k=+0.80$ the armature kinetc energy is 227.4 MJ, so there is a 42\% increase in kinetic energy in the augmented EMG   compared to the non-agumented EMG. 
\begin{table}
\caption{\label{gun_performance}A.C. augmentation.  For parameters in Table~\ref{gun_parameters}, taking the angular frequency of the augmentation circuit voltage in Eq.(\ref{augmentationCircuitVoltage}) as $\omega=10^3$ rad/sec, the  shot time, $t_f$, armature velocity, $v(t_f)$, and armature kinetic energy (KE) in units of mega Joule, are shown for three different values of coupling constant $k$.}
\begin{ruledtabular}
\begin{tabular}{clll} $k$  &  $t_f$ [10$^{-3}$ s]   &   $v(t_f) \,\,$ [10$^3$ m/s]  &  KE $\,\,$ [M J] \\ \hline
 -0.80                       &    5.277     &  3.336                   &  111.3  \\
   0.0                       &    5.545     &  3.994                   &  159.6  \\
  +0.80                      &    5.563     &  4.769                   &  227.4  \\
\end{tabular}
\end{ruledtabular}
\end{table}
\begin{table}
\caption{\label{gun_performance_DC}D.C. augmentation.  For parameters in Table~\ref{gun_parameters}, taking the angular frequency of the augmentation circuit voltage in Eq.(\ref{augmentationCircuitVoltage}) as $\omega=0$, the  shot time, $t_f$, armature velocity, $v(t_f)$, and armature kinetic energy (KE) in units of mega Joule, are shown for three different values of coupling constant $k$.}
\begin{ruledtabular}
\begin{tabular}{clll} $k$  &  $t_f$ [10$^{-3}$ s]   &   $v(t_f) \,\,$ [10$^3$ m/s]  &  KE $\,\,$ [M J] \\ \hline
 -0.80                       &    4.860     &  4.056                   &  164.5  \\
   0.0                       &    5.545     &  3.994                   &  159.6  \\
  +0.80                      &    6.182     &  3.868                   &  149.6  \\
\end{tabular}
\end{ruledtabular}
\end{table}

We can decouple the augmentation circuit from the gun circuit by setting $k=0$, and through Eq.(\ref{MutualInductanceM0}) and (\ref{MutualInductanceMP}), this leads to zero values for mutual the inductances, $M_0=0$ and $M^\prime=0$, in Eq.~(\ref{dynamicEq1})--(\ref{dynamicEq3}).  In this case, Eq.~(\ref{dynamicEq1}) is decoupled from Eq.~(\ref{dynamicEq2}) and Eq.~(\ref{dynamicEq3}). Equation~(\ref{dynamicEq1}) then describes a simple L-R circuit with a voltage generator $V_a(t)$, leading to a current in the (decoupled) augmentation circuit that is purely sinusoidal.  In this case,  Equations~(\ref{dynamicEq2}) and~(\ref{dynamicEq3}) describe a simple EMG that is not coupled to an augmentation circuit. Figure~\ref{fig:k0}  shows the velocity vs. time for this EMG with no augmentation circuit.  Also shown are the currents in the gun circuit and (decoupled) augmentation circuit.  When $k=0$, the time for the armature to reach the end of the rails is $t_f=5.545$ ms and its velocity is $v=3994$ m/s.   For the parameters in Table~\ref{gun_parameters}, the augmented EMG with $k=0.80$  has a 19 \%  increase in velocity over the $k=0$ non-augmented EMG.  As mentioned above, in terms of kinetic energy, the augmented EMG with $k=0.80$ has a 42\% larger kinetic energy than the non-augmented gun (with $k=0.0$), see Figure~\ref{KEplot}, where we plot the kinetic energy of the armature as a function of time.

\begin{figure}[tbp] 
  \centering
   \includegraphics[width=3.5in]{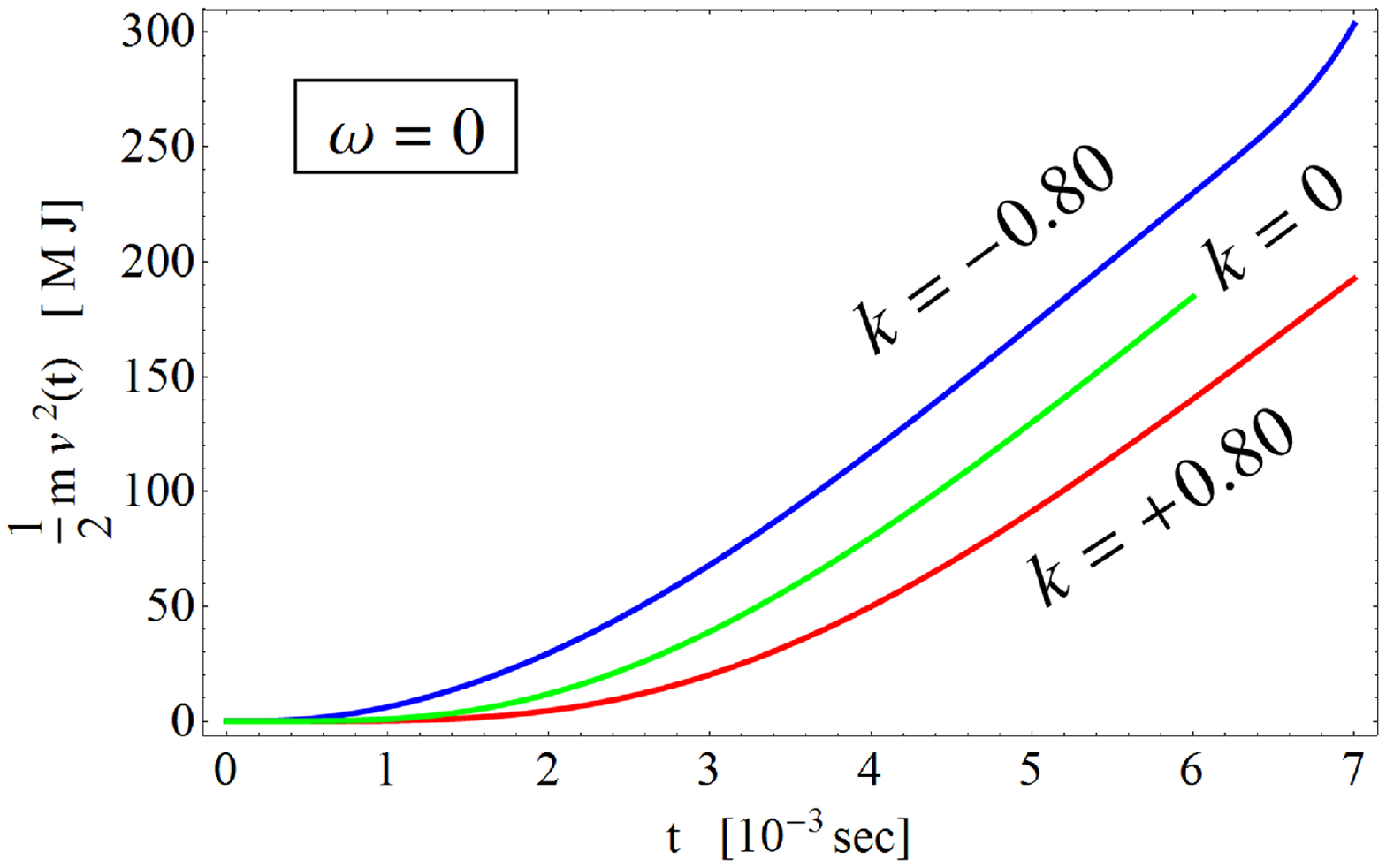}
  \caption{For d.c. augmentation circuit \mbox{voltage (with $\omega=0$),} the armature kinetic energy is plotted vs. time for three values of coupling coefficient, $k=-0.80$, 0.0, and 0.80, for parameters values shown in Table~\ref{gun_parameters}.  The kinetic energy is largest when $k=-0.80$.  For the corresponding velocity plots, see Figure~\ref{fig:w0}.    
  \label{KEplotDC}}
\end{figure}

Finally, we show the effect of the coupling between augmented and gun circuits when the augmented circuit has a d.c. voltage generator with constant amplitude voltage, $V_a(t)= V_{a0}$. We solve Eqs.~(\ref{dynamicEq1})--(\ref{dynamicEq3}) for coupling coefficient values  $k=+0.80$ and $k=-0.80$.  For d.c.  augmentation (with $\omega=0$), for $k=+0.80$, the armature velocity at $t_f=6.182$ ms  is $v=3868$ m/s, while for $k=-0.80$  the armature velocity  at $t_f=4.860$ ms  is $v=4056$ m/s, see  Table~\ref{gun_performance_DC}.  Therefore, for a d.c. voltage generator in the augmentation circuit, the negative value $k=-0.80$ leads to a higher armature velocity than the positive $k$ value, which is just opposite for the case of an a.c. voltage generator in the augmentation circuit.  The plots are shown in Figure~\ref{fig:w0} and the kinetic energy is shown in Figure~\ref{KEplotDC}.  Note that even though the voltage generators in the gun and augmentation circuits are now taken to be d.c., the resulting currents, $I_g(t)$ and $I_a(t)$,  are not constant, due to the  interaction between the augmentation and gun circuits and these circuits interacting with the mechanical degree of freedom $x(t)$ of the armature. 

\section{\label{Summary}Summary}

We have written down the non-linear coupled differential equations for an  augmented EMG that is inductively coupled to the augmentation circuit.   We solved these differential equations numerically using example values of parameters.  For the sample parameters that we used, with $k=+0.80$  the a.c.  augmented EMG had a 42\% larger kinetic energy of the armature than the same EMG with no augmentation.  We have made no effort to vary the parameter values to optimize the augmented EMG performance.  The improvement in performance may be  higher for other parameter values. In this work we have neglected many practical design concerns, such as heating and melting of the rails~\cite{McCorkle2008}. We have considered the simplest configuration of an inductively coupled augmented EMG, with a single external circuit coupling to the EMG circuit.   Other configurations may be explored, such as multi-stage inductively coupled circuits to the EMG circuit.  Further work is needed to explore in detail the parameter space of this augmentation scheme as well as alternative augmentation schemes.

%
\bibliographystyle{apsrev}
\bibliography{References-EMG}
%
\end{document}